\documentclass[12pt]{iopart}

\usepackage{iopams}
\usepackage{color}
\usepackage{amssymb,amsfonts}
\usepackage{graphicx}
\usepackage{bbm}
\usepackage{mathptmx}      

\begin{document}
\title[On the  electric and magnetic properties of conductors]{On the electric and magnetic properties of conductors}

\author{Arbab I. Arbab}
\address{Department of Physics, Faculty of Science, University of Khartoum, P.O. Box 321, Khartoum
11115, Sudan}
\ead{aiarbab@uofk.edu}
\begin{abstract}
Application of the generalized continuity equation reveals that the drift current in conductors is equivalent to a negative diffusion current. A phenomenological model of conductivity is developed using the generalized continuity equations. Consequently, a limiting conductivity is obtained that amounts to $1.09\times \,10^9\,\Omega^{-1}m^{-1}$.  A magnetomotive force (current) is hypothesized to exist, which is exhibited when a voltage changes with time. Magnetic charges and currents are found to be related to displacement current.
\end{abstract}

\maketitle

\section{Introduction}
Diffusion is an important phenomenon that has applications in many disciplines.
Generally, drift fields and/or diffusion processes are the reason for carrier moving velocities. Whereas the drift components  come from an electric drift field, the diffusion parts  originate from carrier density gradients.
In conductors electronic current is transported due to drift current. However, in semi-conductors both diffusion and drift currents contribute to the transport properties. We can treat the flow of electrons inside the material as a flow of fluid. This fluid must satisfy the equation of continuity.
BCS theory is the first microscopic theory of superconductivity, proposed by Bardeen, Cooper, and Schrieffer [1]. It describes superconductivity as a microscopic effect caused by a "condensation" of pairs of electrons into a boson-like state.  Fritz London proposed that the phenomenological London equations may be consequences of the coherence of a quantum state. Superconductivity could be a limiting phenomena for ordinary conductivity of materials. The maximum value of the ordinary conductivity can be limited by some fundamental combination of characteristic constants.
The passage of current should conform with continuity equation. We have recently generalized the ordinary continuity equation to a set of equations [2]. We would like in this paper to apply the generalized continuity equations to study these phenomena, and to explore their consequences. Interestingly, we have found that in conductors, the drift current is equivalent to a diffusion current with negative diffusion constant. Or alternatively, the drift current is a Schrodinger-like current.

\section{The generalized continuity equations}
We have recently generalized the continuity equation to a set of three equations. We would like here to apply this set of equations, aided with Maxwell's equations, to study the electronic properties of materials, viz., conductivity and magnetism. The set of equations reads [2]
\begin{equation}
\vec{\nabla}\cdot \vec{J}+\frac{\partial \rho}{\partial
t}=0\,,
\end{equation}
\begin{equation}
\vec{\nabla}\rho+\frac{1}{c^2}\frac{\partial
\vec{J}}{\partial t}=0\,,
\end{equation}
and
\begin{equation}
\vec{\nabla}\times
\vec{J}=0\,.
\end{equation}
Notice that one can satisfy Eq.(3) by defining a current density by a scalar function $f$ such that
\begin{equation}
\vec{J}=\vec{\nabla}f\,.
\end{equation}
Substituting Eq.(4) in Eq.(2) gives
\begin{equation}
\rho=-\frac{1}{c^2}\frac{\partial f}{\partial t}\,.
\end{equation}
Substituting Eqs.(4) and (5) in Eq.(1)  yields the condition on $f$, viz.,
\begin{equation}
\frac{1}{c^2}\frac{\partial ^2f}{\partial t^2}-\nabla^2 f=0\,.
\end{equation}
Therefore, Eq.(4) is valid for any function that satisfies the wave equation in Eq.(5).
Hence, the current and charge density are not unique. Any redefinition of the type in Eqs.(4) and (5) does not give new physics. This mimics the gauge transformation carried by the vector and scalar potentials $\vec{A}$ and $\varphi$ in electromagnetism. It is interesting to notice here that the same scalar function $f$ that transform  $\vec{A}$ and $\varphi$, i.e.,
\begin{equation}
\vec{A}'=\vec{A}+\vec{\nabla}f\,,\qquad \varphi'=\varphi-\frac{\partial f}{\partial t}\,.
\end{equation}
 transforms
 \begin{equation}
\vec{J}\,'=\vec{J}+\vec{\nabla}f\,,\qquad c^2\rho\,'=c^2\rho-\frac{\partial f}{\partial t}\,,
\end{equation}
where $f$ satisfies Eq.(6).
Thus, the ordinary continuity equation in Eq.(1) is the analogue of the Lorentz gauge in electromagnetism. Hence, the joint transformations in Eq.(7) and (8) make Maxwell's equation invariant. It is interesting to remark that such an essential point has not been considered in formulating Maxwell's equations.

Consider now the integral of Eq.(3) with respect to the surface $S$ and use Stokes theorem to get
\begin{equation}
\int_S\vec{\nabla}\times\vec{J}\cdot d\vec{S}=\int_C\vec{J}\cdot d\vec{\ell}=0\,.
\end{equation}
For a fluid undergoing rigid body flow, the integral
\begin{equation}
\int_C\vec{J}\cdot d\vec{\ell}=2\pi\left(\frac{dL}{dV}\right)=\frac{2\pi L}{V}\,,
\end{equation}
states that the circulation in a fluid is proportional to the angular momentum per unit volume [3]. Since, this circulation vanishes, then the angular momentum of the fluid is conserved.  The vanishing of the circulation implies that the fluid is irrotational. This is known as Kelvin's theorem. Therefore, Eq.(3) represents  angular momentum conservation. It is indeed an amazing result.
The magnetic moment and the angular momentum are related by
\begin{equation}
\mu=-\frac{e}{2m}\, L\,,
\end{equation}
The magnetization is defined as the average magnetic moment per unit volume, i.e.,
\begin{equation}
M=\frac{\bar{\mu}}{V}=-\frac{e}{4\pi\,m}\int_C\vec{J}\cdot d\vec{\ell}\,,
\end{equation}
using Eqs.(10) and (11). Hence, according to Eq.(9), $M=0$.
\section{Conductivity and diffusion properties of fields and currents}
\subsection{Conduction current}
Now consider a conducting media defined by the conduction current density (Ohm's law)
\begin{equation}
\vec{J}=\sigma\,\vec{E}\,.
\end{equation}
Applying Eq.(13) in Eq. (9) yields
\begin{equation}
\int_C\vec{J}\cdot d\vec{\ell}=\sigma\int_C\vec{E}\cdot d\vec{\ell}=\sigma\, V=0\,,\qquad V=\int_C\vec{E}\cdot d\vec{\ell}\,.
\end{equation}
This implies that either $V$ or $\sigma$ must vanish. Since $V\ne0$ then $\sigma=0$. Hence, this equation applies for dielectric materials.

The Maxwell's equations are given by [4]
\begin{equation}
\vec{\nabla}\cdot \vec{E}=\frac{\rho}{\varepsilon_0}\,\,,\qquad\qquad \vec{\nabla}\cdot \vec{B}=0\,,
\end{equation}
and
\begin{equation}
\vec{\nabla}\times \vec{E}+\frac{\partial\vec{B}}{\partial t}=0\,, \qquad \vec{\nabla}\times \vec{B}-\frac{1}{c^2}\frac{\partial\vec{E}}{\partial t}=\mu_0\vec{J}\,.
\end{equation}
Substituting Eq.(13) in Eq.(1) and using Eq.(15), we get
\begin{equation}
\frac{\partial\rho}{\partial t}=-\frac{\sigma}{\varepsilon_0}\,\rho\,.
\end{equation}
Taking the divergence of Eq.(2) and using Eqs. (1), (13) and (15) yield
\begin{equation}
\nabla^2\rho=-\frac{\sigma}{c^2\varepsilon_0}\frac{\partial\rho}{\partial t}\,,\qquad \frac{\partial\rho}{\partial t}=-D_c\nabla^2\rho\,,\qquad D_c=\frac{c^2\varepsilon_0}{\sigma}=\frac{1}{\mu_0\sigma}\,.
\end{equation}
Let us write the charge density is a separable form
\begin{equation}
\rho(r,t)\equiv \rho(r)\rho(t)\,.
\end{equation}
The time dependence of the charge density $\rho(r,t)$ is obtained from Eq.(17) as
\begin{equation}
\rho(t)=\rho_0\exp(-\frac{\sigma}{\varepsilon_0})\,t\,,\qquad \rho_0=\rm const.
\end{equation}
The constant $\tau\equiv\frac{\varepsilon_0}{\sigma}$ is known as the relaxation time, and it is a measure of how fast a conducting medium reaches  electrostatic equilibrium.
The spatial dependence of the charge density is obtained from substituting Eq.(17) in Eq.(18). This yields the equation
\begin{equation}
\nabla^2\rho (r)-\frac{1}{\lambda_A^2}\rho(r)=0\,,\qquad \lambda_A=\frac{\varepsilon_0c}{\sigma}=\frac{1}{\mu_0\sigma c}\,.
\end{equation}
The solution of Eq.(21) is given by
\begin{equation}
\rho(r)=A\,\exp(-\frac{r}{\lambda_A})+A'\exp(\frac{r}{\lambda_A})\,,\qquad A\,\,\,, A'=\rm const.
\end{equation}
Therefore, the charge density distribution (space-time) is given by
\begin{equation}
\rho(r,t)=B\,e^{-\frac{1}{\lambda_A}(\,r+c\,t)}+F\,e^{-\frac{1}{\lambda_A}(-\,r+c\,t)}\,,\qquad B\,, F=\rm const
\end{equation}
This shows that the charge density decays exponentially in space and time.

Now taking the curl of Eq.(3) and employing Eqs.(1), (2) and the vector identity [6]
\begin{equation}
\vec{\nabla}\times(\vec{\nabla}\times \vec{J})=\vec{\nabla}(\vec{\nabla}\cdot \vec{J})-\nabla^2\vec{J}\,,
\end{equation}
  we obtain
\begin{equation}
\frac{\partial \vec{J}}{\partial t}=-D_c\nabla^2\vec{J}\,,\qquad D_c=\frac{1}{\mu_0\,\sigma}\,.
\end{equation}
This equation shows that $\vec{J}$ is governed by Schrodinger-like equation. Hence, both the current density and charged density are governed by a Schrodinger-like equation.

However, Eqs.(2) and (17) suggest that there is a diffusion current (or equivalent to) besides  the drifting current of the form
\begin{equation}
\vec{J}=\left(\frac{c^2\varepsilon_0}{\sigma}\right)\vec{\nabla}\rho=D_c\vec{\nabla}\rho\,.
\end{equation}
This is equivalent to a diffusion current of electrons. It is thought that electrons move in a metal by drifting, but we have seen here the electrical transport properties of metals are propagated by drifting (or equivalently by diffusion) of electrons.
Applying Eq.(26) in Eq.(23), we obtain
\begin{equation}
\vec{J}(r,t)=a\,e^{-\frac{1}{\lambda_A}(\,r+c\,t)}-b\,e^{-\frac{1}{\lambda_A}(-\,r+c\,t)}\,,\qquad a\,, b=\rm const\,.
\end{equation}
The diffusion length is defined by
\begin{equation}
L=\sqrt{D_c\tau}=(R_m\sigma)^{-1}\,,\qquad R_m=\sqrt{\frac{\mu_0}{\varepsilon_0}}\,.
\end{equation}
For silver with $\sigma=6.3\times 10^7\,\Omega^{-1}m^{-1}$, one has $L=6.5\,\mu m$ and $\tau=1.4 \times 10^{-19}\,s$, $D_C=0.0126\, m^2s^{-1}$ and $\lambda_A=0.412\, A^{\rm o}$.

Equations (13) and (25) yield
\begin{equation}
\frac{\partial \vec{E}}{\partial t}=-D_c\nabla^2\vec{E}\,,
\end{equation}
Therefore, $\rho\,,\vec{E}$  and  $\vec{J}$ decay exponentially with
distance and time. Substituting Eq.(13) in the second equation in Eq.(16) yields
\begin{equation}
\vec{\nabla}\times\vec{B}=\mu_0\left(\vec{J}+\frac{1}{\mu_0\sigma\,c^2}\frac{\partial \vec{J}}{\partial t}\right)=\mu_0\left(\vec{J}+\frac{D_c}{c^2}\frac{\partial \vec{J}}{\partial t}\right)\,.
\end{equation}
Upon using Eqs.(2) and (26), Eq.(31) yields
\begin{equation}
\vec{\nabla}\times\vec{B}=0\,.
\end{equation}
Equations (15), (31) and the vector identity $\vec{\nabla}\times(\vec{\nabla}\times\vec{B})=\vec{\nabla}(\vec{\nabla}\cdot\vec{B})-\nabla^2\vec{B}$\,,\, imply that
\begin{equation}
\nabla^2\vec{B}=0\,.
\end{equation}
Equations (31) suggests that   $\vec{B}$ is related to a function $f$ by  $\vec{B}=\vec{\nabla}f$, which implies that $f$  is  harmonic, i.e., $\nabla^2f=0$. This will only work for that part of the magnetic field whose sources are outside of the study region.

Notice when an electromagnetic radiation oscillating with frequency $\omega$ falls on a metal,   the current density is reduced, due to the skin depth ($\delta)$, by [6]
\begin{equation}
J=J_0\exp(-r/\delta)\,,\qquad \delta=\sqrt{\frac{2}{\mu_0\sigma\omega}}\,,
\end{equation}
where $\omega$ is the frequency of the photon.

 There is a theoretical rapid motion of electrons called  \emph{Zitterbewegung } that was first proposed by Schrodinger as a result of his analysis of the wave packet solutions of the Dirac equation for relativistic electrons in free space. For such a case an interference between positive and negative energy states produces what appears to be a fluctuation  of the position of an electron around the median, with a  frequency of $f_c=\frac{2\pi}{\omega_c}=\frac{2mc^2}{h}=\frac{1}{\tau_c}=2.47\times 10^{20}\rm Hz$. This motion is understood to be caused by interference between positive- and negative-energy wave components.

Thus, for such a limiting case, i.e., when $f_c=\frac{2mc^2}{h}=\frac{1}{\tau_c}$, i.e., ultra - high frequency, one finds the combined relation
\begin{equation}
\lambda_c\lambda_A=\delta^2\,.
\end{equation}

Equation (21) may define a limiting conductivity of a material ($\sigma_0$), when we equate $\lambda_A$ to Compton wavelength of the electron.
In this case, one finds
\begin{equation}
\sigma_0=\frac{m}{\mu_0h}\,.
\end{equation}
This amounts to a value of $\sigma_0=1.09\times10^{9}\Omega^{-1}\, m^{-1}$. This can be compared with the best conductivity of metal (Silver) which is $6.63\times 10\,^{7}\Omega^{-1}\, m^{-1}$. Thus, the maximum possible conductivity is set by quantum mechanics and governed by Eq.(35). Because of this the conductivity at zero Kelvin is never infinite but should be limited to $\sigma_0$. Hence, a superconductor should have a non-vanishing resistance that can be limited by Eq.(35). This should be the task of quantum conductivity of metals.
The electrical DC conductivity of metal is given by
\begin{equation}
\sigma=\frac{n\,e^2}{m}\tau\,.
\end{equation}
Equating Eqs.(35) and (36) yields the maximal number density of electrons in the material
\begin{equation}
n_0=\left(\frac{m^2}{\mu_0h\,e^2}\right)\frac{1}{\tau}=\frac{3.88\times 10^{16}}{\tau}\,.
\end{equation}
Since the number densities of electrons in metals are in the range
$10^{28}-10^{29}\,\rm electrons\, m^{-3}$, then the collision time
(or relaxation time), $\tau$ ranges from $10^{-12}-10^{-13}\,\,\rm
sec$. These of course will correspond to the maximum conductivity.

\subsection{Diffusion current}
Diffusion is a transport phenomena resulting from random molecular motion of molecules from a region of higher concentration to a lower concentration. The result of diffusion is a gradual mixing of material.  Diffusion is of fundamental importance in  physics, chemistry, and biology.

The diffusion equation is given by
\begin{equation}
\vec{J}=-D\,\vec{\nabla}\rho\,,
\end{equation}
where  $D$ is the diffusion constant. This is known as Fick's law [6].
Taking the divergence of Eq.(1) and using Eq.(38), one finds
\begin{equation}
\frac{\partial \rho}{\partial t}=D\,\nabla^2 \rho\,.
\end{equation}
This  diffusion equation is equivalent to Eq.(38), hence the density $\rho$ satisfies the diffusion equation. The normalized solution of the Eq.(39) is of the form
\begin{equation}
 \rho (x,t)= \frac{1}{\sqrt{4\pi D\,t}} \exp\,(-\frac{x^2}{4D\,t})\,,
\end{equation}
The diffusion equations, Eqs.(18) and (26), can be compared with drifting  equations, i.e., Eqs.(38) and (39).
Applying Eqs. (38) and (39) in Eq.(2), we obtain
\begin{equation}
\vec{J}-\frac{D}{c^2}\frac{\partial \vec{J}}{\partial t}=0\,.
\end{equation}
This can be compared with Eq.(25), where the diffusion constant $D_c$ replaces $-D$. We, thus, remark that the drift current is equivalent to a diffusion current with negative diffusion constant.
\section{Electric and magnetic properties of conductors}
Now integrate Eq.(2) over the surface $S$ to obtain
\begin{equation}
\int\vec{\nabla}\rho\cdot d\vec{S}+\frac{1}{c^2}\frac{\partial
}{\partial t}\int\vec{J}\cdot d\vec{S}=0\,.
\end{equation}
Using the divergence theorem $\int\vec{A}\cdot d\vec{S}=\int\vec{\nabla}\cdot\vec{A} \,dV\,,$ and Eq.(21), the first term in Eq.(42) yields
\begin{equation}
\int\vec{\nabla}\rho\cdot d\vec{S}=\int\nabla^2\rho\,dV=\frac{1}{\lambda_A^2}\int\rho\,dV=\frac{q}{\lambda_A^2}\,.
\end{equation}
Hence, Eq.(42) becomes
\begin{equation}
\frac{q}{\lambda_A^2}=-\frac{1}{c^2}\frac{\partial I
}{\partial t}\,,\qquad I=\int\vec{J}\cdot d\vec{S}\,.
\end{equation}
This equation implies that
\begin{equation}
-\frac{\partial I
}{\partial t}=\frac{c^2}{\lambda_A^2}\,q=\left(\frac{\sigma}{\varepsilon_0}\right)^2q=\frac{q}{\tau^2}\,.
\end{equation}
This can be written as
\begin{equation}
\frac{d^2q}{d t^2}+\left(\frac{\sigma}{\varepsilon_0}\right)^2q=0\,.
\end{equation}
The solution of the above equation is
\begin{equation}
q(t)=q_{01}\sin(\frac{t}{\tau})+q_{02}\cos(\frac{t}{\tau})\,,\qquad q_{01}\,,\, q_{02}=\rm const\,.
\end{equation}
Hence, the charge oscillates with time.

The voltage ($V$), capacitance ($C$)  and inductance ($L$) are related by
\begin{equation}
V=-L\,\frac{\partial I
}{\partial t}\,,\qquad q=C\,V \,.
\end{equation}
Using  Eq.(45), Eq.(48) yields the relation
\begin{equation}
L=\frac{\tau^2}{C}\,.
\end{equation}
This equation implies that
\begin{equation}
\frac{1}{\sqrt{LC}} =\frac{1}{\tau}=\frac{\sigma}{\varepsilon_0}\equiv f_c\,,
\end{equation}
where $f_c$ is the resonant frequency of LC circuit developed by the material.
Equation (45) can be written as
\begin{equation}
 -\frac{\varepsilon_0}{\sigma^2}\frac{\partial I
}{\partial t}=\frac{q}{\varepsilon_0}\,.
\end{equation}
This equation defines the electric flux, since $\oint\vec{E}\cdot d\vec{S}=\frac{q}{\varepsilon_0}$. Equation (51) suggests that the electric field can be written as
 \begin{equation}
\vec{E}=-\Lambda \frac{\partial \vec{J}}{\partial t}\,,\qquad \Lambda =\frac{\varepsilon_0}{\sigma^2}\,.
\end{equation}
This equation is nothing but the first London equation [5]. However, in London equation $\Lambda$ is some empirical constant.
Consequently, the electric flux can be defined as
\begin{equation}
\varphi_e=-\frac{\partial\Sigma
}{\partial t}\,,\qquad \Sigma=\Lambda\,I\,.
\end{equation}
This equation shows that any change in the current will produce an electric flux. Does the change in the electric flux produce magnetic current in an analogous way to magnetic flux that produces an electric current (or magnetomotive force, $\varepsilon_m$) ? If such a phenomenon exits, we can write
\begin{equation}
\varepsilon_m=-\frac{1}{c}\frac{\partial\varphi_e}{\partial t}=\frac{\varepsilon_0}{\sigma^2c}\frac{\partial^2I}{\partial t^2}=-\frac{C}{\varepsilon_0c}\frac{\partial V}{\partial t}\,.
\end{equation}

It is interesting to notice that Eqs.(52) and (21) can be written as
\begin{equation}
\lambda_A=\sqrt{\frac{\Lambda}{\mu_0}}\,.
\end{equation}
This may coincide with the penetration depth of the magnetic field in a superconducting material.
As the magnetic flux is quantized as [5]
\begin{equation}
\varphi_{0m}=\frac{h}{2e}\,.
\end{equation}
Owing to the symmetry between electricity and magnetism, Eq.(56) suggests that the electric flux passing through any area bounded by  a magnetic current is quantized in units of
\begin{equation}
\varphi_{0e}=\frac{hc}{2e}=6.21\times 10^{-7}\rm V\,m\,.
\end{equation}
Equations (56) and (57) can be written as
\begin{equation}
\varphi_{0e}=c\,\varphi_{0m}\,.
\end{equation}
Alternatively, one may associate a magnetic charge $q_m$ (or
magnetic current, $I_m$) with a quantum electric flux as
\begin{equation}
\varphi_{0e}\,q_m=\frac{h}{2}\,,\qquad or\qquad \varphi_{0m}\,e=\frac{h}{2}\,,
\end{equation}
 so that
\begin{equation}
q_m =e/c=5.33\times 10^{-28}\rm C \,m^{-1} s\,.
\end{equation}
Equation (59) can be thought of an uncertainty relation connecting the two conjugate variables, the magnetic (electric) charge  and electric (magnetic) flux, viz.,
$\Delta\varphi_{e}\,\Delta q_m=\Delta\varphi_{m}\,\Delta e=\frac{h}{2}$.
Thus, the non-detectability of magnetic charge (or magnetic
current) is due to its vanishingly small value only. Magnetic
charges, like quarks, can't be seen as separate particles. The invisibility of quarks is  is a consequence of confinement theory [8].  In similar way, magnetic charges  are confined to move inside magnets. They may have some quantum prescription that does not allow them to appear as singlet states.

Equations (54) yields the \emph{magnetic Ohm's law}
\begin{equation}
\varepsilon_m=\frac{I}{\varepsilon_0c}=I\,R_m\,\qquad I=-C\frac{\partial V}{\partial t}=\frac{\partial q}{\partial t}\,,\qquad
R_m=\frac{1}{\varepsilon_0c}=376.64\,\,\Omega\,.
\end{equation}
It is interesting that $R_m$ is the impedance  of the vacuum. It may also relate to magnetic resistance in magnetic circuits, in an analogous way to electric resistance in electric circuit.
The above relation may usher in the direction that the magnetic charge (current) moves in
vacuum and doesn't require a medium to travel in. The resistance
of the magnetic charge (current) to motion is constant and is
given by the Ohm's law
\begin{equation}
R_m =\frac{1}{\sigma}\frac{L}{A}\,.
\end{equation}
which upon using Eq.(62) can be written as
\begin{equation}
\frac{1}{\varepsilon_0c}
=\frac{1}{\sigma}\frac{L}{A}\qquad\Rightarrow\qquad
\frac{\sigma}{\varepsilon_0c}=\frac{L}{A}\,\qquad\Rightarrow\qquad
\lambda_A=\frac{A}{L}\,.
\end{equation}
Equation (61) implies that
\begin{equation}
\varepsilon_m=-\frac{1}{c}\frac{\partial \varphi_e}{\partial t},\qquad\Rightarrow \qquad\varphi_e=\frac{C\,V}{\varepsilon_0c}=C\, VR_m\,.
\end{equation}
This is equivalent to having an electric flux, $\varphi_e$.

Now integrate Ampere's equation in Eq.(16) to get
\begin{equation}
\int\vec{\nabla}\times \vec{B}\cdot d\vec{S}=\mu_0\int\vec{J}\cdot d\vec{S}+\frac{1}{c^2}\frac{\partial}{\partial t}\int\vec{E}\cdot d\vec{S}\,,
\end{equation}
which yields upon Stokes's law
\begin{equation}
\int \vec{B}\cdot d\vec{\ell}=\mu_0I+\frac{1}{c^2}\frac{\partial\varphi_e}{\partial t}\,.
\end{equation}
Now define the magnetic voltage by
\begin{equation}
V_m=\int c\,\vec{B}\cdot d\vec{\ell}\,,
\end{equation}
so that Eq.(67) yields
\begin{equation}
V_m=\frac{1}{\varepsilon_0c}\left(I+\varepsilon_0\frac{\partial\varphi_e}{\partial
t}\right)=\left(I+\frac{\partial q}{\partial
t}\right)R_m=\left(I+I_D\right)R_m\,.
\end{equation}
Hence, the magnetic voltage, using Eqs.(60) and (61), becomes
\begin{equation}
V_m=\left(I+\frac{1}{\varepsilon_0}\frac{\partial q/c}{\partial
t}\right)R_m=(I+cI_m)R_m\,,\qquad I_m=\frac{\partial q_m}{\partial
t}\,,
\end{equation}
 where $I_m$ is the magnetic current which is related to the displacement current by $I_D=cI_m$. Hence, the displacement current is due to magnetic charge which can pass through the vacuum. This may explain how the direct current (DC) can pass through the capacitor.  The displacement current is associated with the generation of magnetic fields by time-varying electric fields. We see here that magnetic current can flow into vacuum, i.e., open circuit can be closed magnetically. With this enthusiasm, we say that magnetic current flow through a capacitor. Equation (69) represents the magnetic Ohm's law.
\section{The theory of superconductivity-BCS}
BCS theory is the first microscopic theory of superconductivity, proposed by Bardeen, Cooper, and Schrieffer in 1957. It describes superconductivity as a microscopic effect caused by a "condensation" of pairs of electrons into a boson-like state.  Fritz London proposed that the phenomenological London equations may be consequences of the coherence of a quantum state.
If the second of London's equations is manipulated by applying Ampere's law, we obtain
\begin{equation}
\vec{\nabla}\times\vec{B}=\mu_0\vec{J}\,,
\end{equation}
then the result is the differential equation
\begin{equation}
\nabla^2\vec{B}-\frac{1}{\lambda_L^2}\vec{B}=0\,,\qquad \lambda_L=\sqrt{\frac{m}{\mu_0n_se^2}}\,.
\end{equation}
where $n_s$ is the number density of Cooper pairs [5].
Thus, the London equations imply a characteristic length scale, $\lambda_L$, over which external magnetic fields are exponentially suppressed. This value is the London penetration depth.

A simple example geometry is a flat boundary between a superconductor within free space where the magnetic field outside the superconductor is a constant value pointed parallel to the superconducting boundary plane in the $z$ direction. If $x$ leads perpendicular to the boundary then the solution inside the superconductor may be shown to be
\begin{equation}
B_z=B_0\exp(-x/\lambda_L)\,,\qquad B_0=\rm const.\,.
\end{equation}
From here the physical meaning of the London penetration depth can perhaps most easily be discerned. BCS theory correctly predicts the Meissner effect, i.e. the expulsion of a magnetic field from the superconductor and the variation of the penetration depth (the extent of the screening currents flowing below the metal's surface) with temperature. This had been demonstrated experimentally by Walther Meissner and Robert Ochsenfeld [7].

If we now equate $\lambda_A$ to $\lambda_L$\,, we get
\begin{equation}
n_s=\left(\frac{m}{\varepsilon_0e^2}\right)\, \sigma^2\,.
\end{equation}
The maximum value of $n_s$  is reached when $\sigma=\sigma_0$. This yields
\begin{equation}
n_s=\left(\frac{m^3c^2}{\mu_0\, e^2h^2}\right)\, \,.
\end{equation}
This amounts to $n_s=4.8\times 10^{36}\,\,\,\rm \,electrons\,\,/ m^{3}$.
Equation (74) can be written as
\begin{equation}
n_s=\frac{1}{4\pi\,r_c}\frac{1}{\lambda_c^2}\,\,,\qquad n_s=\frac{\alpha^2}{16\pi^3}\frac{1}{r_c^3}\,,\qquad r_c=\frac{e^2}{4\pi\varepsilon_0m\, c^2}\,,
\end{equation}
where $r_c$ is the electron radius. One can also relate the electron radius to the fine structure constant,  $\alpha$, and Compton wavelength by the combined relation, $\alpha=\frac{2\pi\, r_c}{\lambda_c}$. Notice however, that the Thomson cross-section is given by $\sigma_T=\frac{8\pi}{3}\, r_c^2$. Consequently, Eq.(75) can be written as
\begin{equation}
n_s=\frac{\alpha^2}{16}\left(\frac{8}{3\pi}\right)^{3/2}\sigma_T^{-3/2}\,.
\end{equation}
This relation suggests that the motion of free electrons  in a conductor could be associated with some Thomson-like scattering process.
If we now apply Eq.(36) in Eq.(73), we will obtain
\begin{equation}
n_s=\frac{n^2}{n_0}\,\,,\qquad n_0=\frac{\varepsilon_0m}{e^2\tau^2}\,.
\end{equation}
This shows that  $n_s\propto n^2$. Since for metal $\tau\simeq 10^{-14}\rm sec$, $n_0=10^{23}\rm electron/m^3$. Hence, the number of electrons is of the same order of magnitude of Avogadro's number !
\section*{References}
$[1]$ M. Tinkham, \emph{Introduction to Superconductivity}, MacGraw-Hill, Inc., New York (1996).\\
$[2]$ Arbab, A. I., and Widatallah, H. M., \emph{The generalized continuity equations}, http://arxiv.org/abs/1003.0071.\\
$[3]$  Dolzhanskii, F. V., \emph{Mechanical prototypes of fundamental hydrodynamic invariants}, \emph{Izvestiya, Atmospheric and Oceanic Physics}, \textbf{37}, no.4, 2001 (2001). (Alternatively you may look at: http://maxwell.ucdavis.edu/~cole/phy9b/notes/$fluids_-$ch3.pdf)\\
$[4]$ David J. Griffiths., \emph{Introduction to Electrodynamics}, 3rd ed., Addison Wesley, (1999).\\
$[5]$ London, F.; H. London , "The Electromagnetic Equations of the Supraconductor", \emph{Proc. Roy. Soc. (London)}, \textbf{A149}, (866), 71 (1935). \\
$[6]$ Fick A., \emph{Ann. Physik, Leipzig}, \textbf{170}, 59, (1855).\\
$[7]$ Meissner, W., and R. Ochsenfeld.,  \emph{Ein neuer effekt bei eintritt der supraleitfahigkeit}, \emph{Naturwissenschaften} \textbf{21}, 44, 787 (1933).\\
$[8]$  Barger,V. and  Phillips,  R.,  \emph{Collider Physics}, Addison–Wesley, (1997); Wilson, K. G., et al., \emph{Phys Rev D}\textbf{49}, pp 6720 (1994).\\
\end{document}